\documentstyle[12pt,amsmath,amsfonts, graphicx, hyperref]{article}
\newcommand{\lC}{\mathrm{l\hspace{-2.1mm}C}}
\newcommand{\lR}{\mathrm{I\hspace{-0.7mm}R}}

\numberwithin{equation}{section}

\begin{document}

\hoffset = -1truecm \voffset = -2truecm

\begin{titlepage}
{\flushright \today
\\}
\vskip 2truecm {\center \LARGE \bf $N=1$ Supergravity BPS Domain
Walls on K\"ahler-Ricci Soliton
\\} \vskip 2truecm

{\center \bf
 Bobby E. Gunara$^{\flat}$, Freddy P. Zen$^{\flat}$, and Arianto$^{\flat,\sharp}$
 \footnote{email: bobby@fi.itb.ac.id, fpzen@fi.itb.ac.id, feranie@upi.edu}
\\}

\vskip 1truecm

\begin{center}
$^{\flat}$\textit{Indonesia Center for Theoretical and
Mathematical Physics (ICTMP),
\\
and \\
Theoretical  Physics Laboratory, \\
Theoretical High Energy Physics and Instrumentation Research
Group,\\
Faculty of Mathematics and Natural Sciences, \\
Institut Teknologi Bandung \\
Jl. Ganesha 10 Bandung 40132, Indonesia.}

 \vskip 0.5truecm

$^{\sharp}$\textit{Department of Physics, \\
Faculty of Mathematics and Natural Sciences, \\
Udayana University \\
Jl. Kampus Bukit Jimbaran-Kuta Denpasar 80361, Indonesia.}

\end{center}

\vskip 2truecm

{\center \large \bf ABSTRACT
\\}
\vskip 0.5truecm

\noindent This paper provides a study of some aspects of flat and
curved BPS domain walls together with their Lorentz invariant
vacua of four dimensional chiral $N=1$ supergravity. The scalar
manifold can be viewed as a one-parameter family of K\"ahler
manifolds generated by a K\"ahler-Ricci flow equation.
Consequently, a vacuum manifold characterized by $(m,\lambda)$
where $m$ and $\lambda$ are the dimension and the index of the
manifold, respectively, does deform with respect to the flow
parameter related to the geometric soliton. Moreover, one has to
carry out the renormalization group analysis to verify the
existence of such a vacuum manifold in the ultraviolet or infrared
regions. At the end, we discuss a simple model with linear
superpotential on $U(n)$ symmetric K\"ahler-Ricci orbifolds.

\vskip 0.5truecm


\end{titlepage}




\section{Introduction}
A K\"ahler-Ricci flow equation has gained both mathematical and
physical interests due to several developments. In the
mathematical context, particularly in compact K\"ahler manifolds
with the first Chern class $c_1 =0$ or $c_1 <
 0$, this equation could give a new proof of the well-known Calabi
conjecture \cite{cao0}. Moreover, it has been shown the existence
of solutions of the K\"ahler-Ricci equation with $U(n)$ symmetry
on $\, {\lC}^n$ \cite{cao, cao1} and line bundles over
$\,{\mathrm{\lC P}}^{n -1}$ \cite{FIK}.\\
\indent In the physical context, for example, four dimensional
chiral $N=1$ supersymmetries on K\"ahler-Ricci solitons generating
a one-parameter family of K\"ahler manifolds together with their
domain wall solutions have been studied by two of the authors in
serial papers \cite{GZ, GZ1, GZ2}. Those works might provide
preliminary studies of the Anti de Sitter/Conformal Field Theory
(AdS/CFT) correspondence (for a review, see for example
\cite{AdSCFT}) in four dimensions and its evolution on special
cases of K\"ahler-Ricci solitons, namely K\"ahler-Einstein
manifolds, $U(n)$ symmetric cone-dominated K\"ahler-Ricci solitons
on line bundles over $\,{\mathrm{\lC P}}^{n -1}$ \cite{FIK}, and
two
dimensional K\"ahler-Ricci solitons.\\
 \indent The aim of this paper is to
study flat and curved BPS domain walls of four dimensional $N=1$
supergravity coupled to chiral multiplets whose scalar manifold is
considered to be a one-parameter family of K\"ahler geometries
generated by the K\"ahler-Ricci flow equation. This soliton can be
regarded as a volume deformation of a K\"ahler geometry for finite
$\tau \in \lR$, where $\tau$ is the flow parameter. Two direct
consequences of such treatment can be mentioned as follows.
Firstly, all couplings such as the shifting quantities, the masses
of the fields, and the scalar potential evolve with respect to
$\tau$ since those quantities depend on the shape of geometries
\cite{GZ, GZ1, GZ2}. Secondly, it is possible to have a situation
where the signature of the metric changes. For example, as
observed in our previous works \cite{GZ, GZ1, GZ2} a model with
positive definite metric could be changed into a model with
negative definite metric. At the classical level such a situation
is permitted, but at the quantum level this would provides a
negative norm states in the Fock space, which are removed as
unphysical \footnote{See also Section \ref{SUGRA}.}.\\
\indent Then, several aspects of their Lorentz invariant vacua
will also be discussed. Our study here is in the context of
Morse(-Bott) theory and the renormalization group (RG) flow
analysis that generalizes the previous works for flat domain walls
on K\"ahler-Einstein manifolds and the $U(n)$ symmetric
cone-dominated K\"ahler-Ricci solitons \cite{GZ}, and also, curved
(AdS sliced) domain walls on two dimensional K\"ahler-Ricci
solitons \cite{GZ1}. At the end, we apply the study to a case of
$U(n)$ symmetric K\"ahler-Ricci solitons admitting an
orbifold-type singularity at the origin
(called $U(n)$ symmetric K\"ahler-Ricci orbifolds) \cite{FIK}.\\
 \indent We organize this paper as
follows. Section \ref{SKRF} is devoted to review general
descriptions of K\"ahler-Ricci soliton and the construction of an
example, namely  $U(n)$ symmetric K\"ahler-Ricci orbifolds. Then,
some possible cases of the soliton near the orbifold point (the
origin) which can be viewed as a one-parameter family of K\"ahler
manifolds are summarized in Theorem \ref{Korbi} and Lemma
\ref{Korbi1}. In Section \ref{SUGRA} we provide a review of $N=1$
supergravity coupled to chiral multiplets where the scalar
manifold is considered to be a one-parameter family of K\"ahler
manifolds generated by a K\"ahler-Ricci soliton. In Section
\ref{flatDW} flat domain walls are reviewed and then, we study
some aspects of their vacuum structure on general K\"ahler-Ricci
soliton using Morse(-Bott) theory and the RG flow analysis. Some
results are written down in Theorem \ref{LemmaVmorse} and Theorem
\ref{Modmorselemma}. Then, we extend the previous case to curved
domain walls in Section \ref{curvedDW}. In Section \ref{VSNO} we
discuss properties of vacua of $N=1$ theory on $U(n)$ symmetric
K\"ahler-Ricci orbifolds near the origin. Finally, we put our
conclusions in Section \ref{conclu}.


\section{K\"ahler-Ricci Flow}
\label{SKRF}
 The structure of this section is as follows. Firstly,
we review some general aspects of the the K\"ahler-Ricci flow.
Secondly, the construction of K\"ahler-Ricci solitons on complex
line bundles over $\,{\mathrm{\lC P}}^{n -1}\; , n \ge 2, $ with
orbifold-type singularities called K\"ahler-Ricci orbifolds will
also be reviewed \cite{FIK}. Then, we particularly discuss some
possible cases where the soliton near the orbifold point can be
viewed as a one-parameter family of K\"ahler manifolds. Here, our
convention follows rather closely the reference \cite{GZ}
\footnote{We also recommend \cite{topping, chow, caozhu} for a
review of the Ricci flow equation.}.

\subsection{General Picture}

\indent Let us first consider a general picture of the
K\"ahler-Ricci soliton. A complex K\"ahler manifold
$({\bf{M}},g(\tau))$ satisfying
\begin{equation}
\frac{\partial g_{i\bar{j}}}{\partial \tau}(z, \bar{z}; \tau) = -2
R_{i\bar{j}}(z, \bar{z}; \tau)\;, \quad 0 \le \tau < T \;,
\label{KRF}
\end{equation}
is called K\"ahler-Ricci soliton where $(z, \bar{z}) \in {\bf{M}}$
and the metric
\begin{equation}
g(\tau) \equiv g_{i\bar{j}}(z, \bar{z}; \tau) \; dz^i
d\bar{z}^{\bar{j}} \;.
\end{equation}
In particular, the metric $g(\tau)$ can be written as
\begin{equation}
g(\tau) = \sigma(\tau) \psi^*_{\tau}(g(0))\;, \quad 0 \le \tau < T
\;, \label{KRsolumum}
\end{equation}
with $\sigma(\tau) \equiv (1-2\Lambda \tau)$ and the
diffeomorphism map $\psi_{\tau}$. The initial metric at $\tau =0$,
namely $g(0)$, fulfills the following relation
\begin{equation}
-2R_{i\bar{j}}(0) = \nabla_i Y_{\bar{j}}(0)+
\bar{\nabla}_{\bar{j}} Y_i(0) -2\Lambda
g_{i\bar{j}}(0)\;,\label{initialgeom}
\end{equation}
 for $\Lambda \in \lR$ and some holomorphic vector
fields \footnote{Holomorphicity of $Y(0)$ is coming from the fact
that we impose the condition $\psi^*_{\tau}(J)= J$ on ${\bf{M}}$
where $J$ is the complex structure on ${\bf{M}}$.}
\begin{equation}
Y(0) = Y^i(z, 0) \partial_i + Y^{\bar{i}}(\bar{z}, 0)
\bar{\partial}_{\bar{i}}\;, \label{killdiff0}
\end{equation}
 on ${\bf{M}}$ where $i,j = 1,...,
 {\text{dim}}_{\:\mathrm{l\hspace{-1.6mm}C}}({\bf{M}})$.
 Moreover, using the vector field
$Y(0)$, we can define a $\tau$-dependent vector field $X(\tau)$
\begin{equation}
X(\tau) = \frac{1}{\sigma(\tau)}Y( 0)\;, \label{killdiff}
\end{equation}
generating a family of diffeomorphisms $\psi_{\tau}$. In addition,
the vector field $X(\tau)$ satisfies the following equation
\begin{eqnarray}
\frac{\partial \hat{z}^i}{\partial \tau} &=& X^i(\hat{z}, \tau)
\;,\label{diffeo}
\end{eqnarray}
where
\begin{equation}
\hat{z} \equiv \psi_{\tau}(z) \;. \label{diffeo1}
\end{equation}

\subsection{An Example: $U(n)$ Symmetric K\"ahler-Ricci Orbifold}

\indent Before constructing K\"ahler-Ricci orbifolds, we firstly
discuss the construction of a K\"ahler metric on $ \,{\lC}^n
\backslash \{0 \}$. Our starting point is to define the initial
K\"ahler potential at $\tau =0$, namely
\begin{equation}
K(z, \bar{z},0) =  \phi(u) \;, \label{Kpotans}
\end{equation}
where
\begin{equation}
u \equiv 2\,{\mathrm{ln}}(\delta_{i\bar{j}}z^i \bar{z}^{\bar{j}})
= 2\,{\mathrm{ln}} \vert z \vert^2\;, \label{newu}
\end{equation}
on $ \,{\lC}^n \backslash \{0 \}$. Thus, the ansatz
(\ref{Kpotans}) maintains a $U(n)$ symmetry. For a sake of
simplicity we take the vector field $Y^i(0)$ to be holomorphic and
linear
\begin{equation}
Y^i(0) =  \mu z^i \;, \label{killvek}
\end{equation}
where $\mu \in \lR$. Then, the K\"ahler potential (\ref{Kpotans})
implies that the metric is given by
\begin{equation}
g(0) = g_{i\bar{j}}(0)\, dz^i \, d\bar{z}^{\bar{j}} = \left[ 2
e^{-u/2}\phi_u \,\delta_{i\bar{j}} + 4
e^{-u}(\phi_{uu}-\frac{1}{2}\phi_u) \,\bar{z}^{\bar{i}} \, z^j
\right] dz^i \, d\bar{z}^{\bar{j}} \;,\label{metricans}
\end{equation}
together with its inverse
\begin{equation}
g^{-1}(0) = g^{\bar{j}i}(0)\, \bar{\partial}_{\bar{j}}\,
\partial_i = \frac{e^{u/2}}{2\phi_u}
\left[
\delta^{\bar{j}i}-e^{-u/2}\,\frac{\phi_{uu}-\frac{1}{2}\phi_u}{\phi_{uu}}\,
\bar{z}^{\bar{j}} z^i \right]\, \bar{\partial}_{\bar{j}}\,
\partial_i  \;,\label{inversmetricans}
\end{equation}
 where the positivity of the metric (\ref{metricans}) implies
\begin{equation}
\phi_u \equiv \frac{d\phi}{du} > 0 \;, \quad \phi_{uu} \equiv
\frac{d^2\phi}{du^2} > 0 \;. \label{inequal}
\end{equation}
 Inserting (\ref{killvek}) and (\ref{metricans}) into
 (\ref{initialgeom}), and defining $\Phi \equiv \phi_u$
 and  $\Phi_u \equiv F(\Phi)$, we then obtain
\begin{equation}
\frac{dF}{d\Phi}+(\frac{n -1}{\Phi}-4\mu)F - (\frac{n
}{2}-2\Lambda \Phi) = \frac{A_0}{\Phi} e^{(1-n)u/2}\;,
\label{FPhi}
\end{equation}
with $A_0$ is an arbitrary constant. Taking simply $A_0 =0$, the
solution of (\ref{FPhi}) has the form
\begin{equation}
\Phi_u = F(\Phi) = A_1 \frac{e^{4\mu\Phi}}{\Phi^{n -1}} +
\frac{\Lambda}{2\mu}\Phi  +  \frac{2(\Lambda -\mu)}{(4\mu)^{n
+1}}\sum_{j=0}^{n -1} \frac{n!}{j!} (4\mu)^j \,\Phi^{j+1-n} \;,
\label{solFPhi}
\end{equation}
where $A_1$ is also an arbitrary constant. Note that for $\mu =0$
we obtain the K\"ahler-Einstein geometry as the solution of (\ref{FPhi}).\\
\indent Recalling (\ref{killdiff}) and (\ref{killvek}) we find
that the vector field $X^i(\tau)$ has the form
\begin{equation}
X^i(\tau) =  \frac{\mu}{\sigma(\tau)} z^i \;, \label{killdifvek}
\end{equation}
which generates the diffeomorphisms
\begin{equation}
\hat{z} \equiv \psi_{\tau}(z) = \sigma(\tau)^{-\mu/2\Lambda}z \;.
\label{diffeo2}
\end{equation}
Thus, the complete $\tau$-dependent K\"ahler-Ricci soliton on  $
\,{\lC}^n \backslash \{0 \}$ is given by
\begin{equation}
 g(z,\bar{z};\tau) = \sigma(\tau)^{1-\mu/\Lambda}
 g_{i\bar{j}}\left(\sigma(\tau)^{-\mu/2\Lambda}z \right)
 dz^i d\bar{z}^{\bar{j}}\;. \label{KRsol}
\end{equation}
\indent Now, we turn to construct the gradient K\"ahler-Ricci
soliton on line bundles over $\,{\mathrm{\lC P}}^{n -1}$ with $n
\ge 2$. Let us first define the metric (\ref{metricans}) on $(
\,{\lC}^n \backslash \{0 \})/{\mathbb{Z}}_{\ell}$, where
${\mathbb{Z}}_{\ell}$ acting on $\,{\lC}^n \backslash \{0 \}$ by
$z  \mapsto e^{2\pi {\mathrm{i}}/\ell}z$ with $\ell$ is a positive
integer and $\ell \ne 0$. A positive line bundle over
$\,{\mathrm{\lC P}}^{n -1}$,
 denoting by $L^{\ell}$, can then be constructed by gluing $\,{\mathrm{\lC
P}}^{n -1}$ into $( \,{\lC}^n \backslash \{0
\})/{\mathbb{Z}}_{\ell}$ at infinity. Note that for $\ell =1$, $L$
is called hyperplane bundle which is dual to the tautological line
bundle $L^{-1}$. Moreover, we replace the above coordinates $z$ by
new coordinates of the form
\begin{equation}
\xi^i \equiv (z^i)^{\ell} \;,\label{newcoor}
\end{equation}
 parameterizing $(\,{\lC}^n \backslash \{0
 \})/{\mathbb{Z}}_{\ell}$. This follows that around infinity,
 the initial metric (\ref{metricans}) can be changed into the form
\begin{equation}
 g(0) = \Phi \, g_{FS} + \Phi_v \, dw\,d\bar{w}\;,
\end{equation}
where $g_{FS}$ is the standard Fubini-Study metric of
$\,{\mathrm{\lC P}}^{n -1}$
\begin{equation}
g_{FS} = \left( \frac{\delta_{a\bar{b}}}{1+\zeta^c
\bar{\zeta}^{\bar{c}}}-\frac{\zeta^b
\bar{\zeta}^{\bar{a}}}{(1+\zeta^c
\bar{\zeta}^{\bar{c}})^2}\right)d\zeta^a
\,d\bar{\zeta}^{\bar{b}}\;, \label{FSmetric}
\end{equation}
 with $\zeta^a \equiv \xi^a /\xi^n$, $\zeta^n =1$ and $a,b,c = 1,...,n -1$.
 Here, $w \equiv w(\xi,\bar{\xi})$ is a
nonholomorphic coordinate and $v \equiv 2\,{\mathrm{ln}} \vert \xi
\vert^2$. So, in this case we have
\begin{equation}
 \lim_{v \to +\infty} \Phi(v) = a > 0 \;, \quad F(a)= 0 \;,
 \quad \frac{dF}{d\Phi}(a) < 0 \;, \label{inftycon}
\end{equation}
with
\begin{equation}
a = \frac{1}{4 \Lambda } (n + \ell) \;.
\end{equation}
The condition $F(a) = 0$ implies
\begin{equation}
  A_1(\Lambda, a; \mu) = -e^{4\mu a} \left\lbrack
  \frac{\Lambda}{2\mu} \, a^n
  + \frac{2 n! (\Lambda - \mu)}{(4\mu)^{n+1}} \sum_{j=0}^{n-1}
  \frac{(4 \mu )^j}{j!} \, a^j \right\rbrack \;.
   \label{zeroF}
\end{equation}
For the case at hand we only have $\Lambda
> 0$ and $ \ell > 0$ which describe a shrinking K\"ahler-Ricci
soliton.\\
\indent Next, let us consider a condition for adding a point to $(
\,{\lC}^n \backslash \{ 0 \})/{\mathbb{Z}}_{\ell}$ at $z=0$. We
begin by mentioning a similar condition as (\ref{inftycon}),
namely
\begin{equation}
 \lim_{v \to -\infty} \Phi(v) =  0 \;, \quad F(0)= 0 \;,
 \quad \frac{dF}{d\Phi}(0) > 0 \;. \label{addzerocon}
\end{equation}
From $F(0)= 0$, it follows
\begin{equation}
  A_1(\Lambda, 0; \mu) = -\frac{2 n! (\Lambda - \mu)}{(4\mu)^{n+1}}  \;.
   \label{zero0}
\end{equation}
After some computations, we then obtain
\begin{equation}
  \Phi(v) = \vert \xi \vert^2 B(\vert \xi \vert^2)
  \;, \label{nearzerocon}
\end{equation}
where $B(0) > 0$. It is easy to see that around $z=0$, the initial
metric (\ref{metricans}) simplifies to
\begin{equation}
g(0) =   2 \Big\lbrack B\!\left(\vert \xi \vert^2 \right)
 \,  \delta_{i\bar{j}} + \dot{B}
 \! \left(\vert \xi \vert^2 \right)
    \, \bar{\xi}^{\bar{i}}
 \xi^j \Big\rbrack \, d\xi^i d\bar{\xi}^{\bar{j}}
  \;. \label{nearzerometric}
\end{equation}
The complete $\tau$-dependent soliton has the form
\begin{equation}
  g(\tau) = 2 \, \sigma(\tau)^{1-\mu /\Lambda} \Bigg\lbrack B
  \! \left(\sigma(\tau)^{-\mu /\Lambda}\vert \xi \vert^2 \right)
 \,  \delta_{i\bar{j}} + \sigma(\tau)^{-\mu /\Lambda} \dot{B}
  \! \left(\sigma(\tau)^{-\mu /\Lambda}\vert \xi \vert^2 \right)
   \, \bar{\xi}^{\bar{i}}
 \xi^j \Bigg\rbrack \, d\xi^i d\bar{\xi}^{\bar{j}}
  \;. \label{nearzerosoliton}
\end{equation}

 \indent Some comments are as follows. Firstly, taking
(\ref{zeroF}) and (\ref{zero0}) satisfy simultaneously, we find
that for $\Lambda \ge 1$ and small $\mu >0$, one positive root
$\mu$ indeed exists in $0 < \mu < \Lambda$. Secondly, the
condition (\ref{addzerocon}) means $L^{\ell} \cup \{0 \} \simeq
\,{\mathrm{\lC P}}^n /{\mathbb{Z}}_{\ell}$. For $\ell =1$, we have
$\mu =0$ which means that the soliton is the K\"ahler-Einstein
geometry $\,{\mathrm{\lC P}}^n$. In other words, we smoothly add a
point at $z=0$ and the initial metric (\ref{nearzerometric})
becomes flat. In the other case, namely the $\ell \ge 2$ case, the
metric (\ref{nearzerometric}) vanishes at $z=0$ showing that this
is an orbifold singularity at the origin. Finally, the flow
(\ref{nearzerosoliton}) also diverges at
$\tau = 1/2\Lambda$.\\
\indent For the rest of the paper we will consider the case where
the K\"ahler-Ricci soliton can be viewed as one-parameter family
of K\"ahler manifolds. In order to obtain a consistent picture we
take the functions $B\!\left(\sigma(\tau)^{-\mu /\Lambda} \vert
\xi \vert^2 \right)$ and $\dot{B} \! \left(\sigma(\tau)^{-\mu
/\Lambda}\vert \xi \vert^2 \right)$ to be real valued for $\tau
\ge 0$ and $\tau \ne 1/2\Lambda$. For the case at hand, namely the
soliton (\ref{nearzerosoliton}), the first possibility can be
written down in the following statements.

\newtheorem{Korbi}{Theorem}
\begin{Korbi}\label{Korbi}
Let us consider the K\"ahler-Ricci soliton around the origin
(\ref{nearzerosoliton}) where $ p/q \equiv 1-\mu /\Lambda $ is
taken to be a rational number with $0 < p <q$. Assuming that there
exist such $p$ and $q$ for $\Lambda \ge 1$ and $0 < \mu <
\Lambda$. Then, the possible cases are as follows.
\begin{itemize}
\item[1.] If both $p$ and $q$ are odd integers, one has then
\begin{eqnarray}
\sigma(\tau)^{p/q} > 0 &\quad& \mathrm{for} \: \: \tau < 1/2\Lambda
\;,\nonumber\\
\sigma(\tau)^{p/q} < 0 &\quad& \mathrm{for} \: \: \tau >
1/2\Lambda \;. \label{specase}
\end{eqnarray}
The functions $B\!\left(\sigma(\tau)^{-\mu /\Lambda} \vert \xi
\vert^2 \right)$ and $\dot{B} \! \left(\sigma(\tau)^{-\mu
/\Lambda}\vert \xi \vert^2 \right)$ are positive definite for
$\tau \ge 0$ and $\tau \ne 1/2\Lambda$.

\item[2.] If $p$ and $q$ are even and odd integers respectively,
then $\sigma(\tau)^{p/q} > 0$ for $\tau \ge 0$ and $\tau \ne
1/2\Lambda$. But, the sign of both $B\!\left(\sigma(\tau)^{-\mu
/\Lambda} \vert \xi \vert^2 \right)$ and $\dot{B} \!
\left(\sigma(\tau)^{-\mu /\Lambda}\vert \xi \vert^2 \right)$ might
be altered after hitting the singularity at $\tau = 1/2\Lambda$.

\item[3.] If $p$ and $q$ are odd and even respectively, then
$\sigma(\tau)^{p/q}$ becomes imaginary for $\tau
> 1/2\Lambda$. In other words, no K\"ahler manifold exists for  $\tau
> 1/2\Lambda$.
\end{itemize}
\end{Korbi}

The second possibility is then

\newtheorem{Korbi1}[Korbi]{Lemma}
\begin{Korbi1}\label{Korbi1}
If $\, 1-\mu /\Lambda $ is an irrational number, then it does not
exist any K\"ahler manifold for  $\tau> 1/2\Lambda$ because
$\sigma(\tau)^{1-\mu /\Lambda }$ turns into a complex number for
$\tau > 1/2\Lambda$.
\end{Korbi1}

We put the proof of Theorem \ref{Korbi} and Lemma \ref{Korbi1} in
Appendix \ref{PT2}. In addition, since $\xi \approx 0$ we could
have
\begin{eqnarray}
B\!\left(\sigma(\tau)^{-\mu /\Lambda} \vert \xi \vert^2 \right)
&\approx& B(0) \;,\nonumber\\
\dot{B} \! \left(\sigma(\tau)^{-\mu /\Lambda}\vert \xi \vert^2
\right) &\approx& \dot{B}(0) \;,
\end{eqnarray}
for some finite $\tau$ that far away from $\tau = 1/2\Lambda$.
This is the simplest case of Theorem \ref{Korbi} and Lemma
\ref{Korbi1} where the functions $B$ and $\dot{B}$ are positive
definite. Some of these results will be applied to study a vacuum
structure of the $N=1$ theory in Section \ref{VSNO}.

\section{$N=1$ Chiral Supergravity on K\"ahler-Ricci Soliton}
\label{SUGRA}
 This section is devoted to review properties of four
dimensional $N=1$ chiral supergravity on a one-parameter family of
K\"ahler manifolds generated by a K\"ahler-Ricci soliton
\cite{GZ}. Here, we only consider some aspects which are useful
for our analysis for the rest of the paper. Some excellent
references for a review of $N=1$ supergravity in four dimensions
can be found, for example
in \cite{susy, DF}.\\
 \indent The $N=1$ theory is simply a
gravitational multiplet coupled with $n$ chiral multiplets. The
gravitational multiplet has a vierbein $e^a_{\mu}$ and a vector
spinor $\psi_{\mu}$ where $a=0,...,3$ and $\nu=0,...,3$ are the
flat and the curved indices, respectively. The chiral multiplet
consists of a complex scalar $z$ and a spin-$\frac{1}{2}$ fermion
$\chi$.\\
\indent The construction of the local $N=1$ theory on a
K\"ahler-Ricci is as follows. First, we consider the Lagrangian in
\cite{susyor} as the initial Lagrangian at $\tau = 0$. Then, by
replacing all couplings that depend on the geometric quantities
such as the metric $g_{i\bar{j}}(0)$ by the soliton
$g^{i\bar{j}}(\tau)$, the bosonic parts of the Lagrangian has the
form
\begin{equation}
{\mathcal{L}}^{N=1} = -\frac{M^2_P}{2}R + g_{i\bar{j}}(z, \bar{z};
\tau)\,
\partial_{\nu} z^i \,
\partial^{\nu}\bar{z}^{\bar{j}} - V(z,\bar{z}; \tau)\;, \label{L}
\end{equation}
where $M_P$ is the Planck mass. The quantity $R$ is the Ricci
scalar of the four dimensional spacetime; the scalar fields
$(z,\bar{z})$ span a Hodge-K\"ahler manifold endowed with metric
$g_{i\bar{j}}(z, \bar{z}; \tau) \equiv
\partial_i
\partial_{\bar{j}}K(z,\bar{z}; \tau)$ satisfying (\ref{KRF});
and $K(z,\bar{z}; \tau)$ is a real function, called the K\"ahler
potential. Then, the $N=1$ scalar potential $V(z,\bar{z}; \tau)$
has the form
\begin{equation}
 V(z,\bar{z}; \tau) = e^{K(\tau)/M^2_P}\left(g^{i\bar{j}}(\tau)\nabla_i W\,
 \bar{\nabla}_{\bar{j}} \bar{W}
 - \frac{3}{M^2_P} W \bar{W} \right)\;,
\label{V01}
\end{equation}
where $W$ is a holomorphic superpotential and $\nabla_i W\equiv
\partial_i W + (K_i(\tau)/M^2_P) W$. Furthermore, the Lagrangian
(\ref{L}) admits a symmetry realized in the following
supersymmetry transformations up to three-fermion terms
\begin{eqnarray}
\delta\psi_{1\nu} &=& M_P \left(D_{\nu}\epsilon_1 +
\frac{\mathrm{i}}{2}e^{K(\tau)/2M^2_P}\,W \gamma_{\nu} \epsilon^1
+ \frac{\mathrm{i}}{2M_P} Q_{\nu}(\tau)\epsilon_1 \right)\;, \nonumber\\
\delta\chi^i &=& {\mathrm{i}}\partial_{\nu} z^i \, \gamma^{\nu}
\epsilon^1
+ N^i(\tau) \epsilon_1 \quad, \label{susytr}\\
\delta e^a_{\nu} &=&  - \frac{{\mathrm{i}}}{M_P} \, (
\bar{\psi}_{1\nu} \, \gamma^a \epsilon^1
+ \bar{\psi}^1_\nu \, \gamma^a \epsilon_1 )\;,\nonumber\\
\delta z^i &=& \bar{\chi}^i \epsilon_1 \;,\nonumber
\end{eqnarray}
where $N^i(\tau) \equiv
e^{K(\tau)/2M^2_P}\,g^{i\bar{j}}(\tau)\bar{\nabla}_{\bar{j}}\bar{W}$,
 $g^{i\bar{j}}(\tau)$ is the inverse of $g_{i\bar{j}}(\tau)$,
 and the $U(1)$ connection $Q_{\nu}(\tau) \equiv  - \left(  K_i(\tau)
 \,\partial_{\nu}z^i -
K_{\bar{i}}(\tau)\, \partial_{\nu}\bar{z}^{\bar{i}}\right)$. In
addition, we have also introduced $\epsilon_1 \equiv
\epsilon_1(x,\tau)$.\\
 \indent It is important to notice, as
mentioned in the previous section, that the metric
$g_{i\bar{j}}(\tau)$ could possibly turn into a negative definite
metric. Such a theory would be a non-unitary with ghosts. At the
classical level, it may be possible to have theories of this type.
However, at the quantum level this would provide negative norm
states in the standard Fock space which are discarded as
unphysical \cite{GZ2}.

\section{Flat BPS Domain Walls on K\"ahler-Ricci Soliton}
\label{flatDW}
 The organization of this section is as follows.
First, we provide a short review of flat BPS domain walls of the
four dimensional $N=1$ supergravity coupled to arbitrary chiral
multiplets whose the scalar fields span a one-parameter family of
K\"ahler manifolds generated by the K\"ahler-Ricci flow
(\ref{KRF}). These have originally discussed in \cite{GZ}
\footnote{Flat BPS domain walls have been firstly considered in
\cite{FlatDW} and reviewed in \cite{RevFlatDW}.}. Here, we use
similar convention as in \cite{GZ, CDGKL, GZA}. Then, we describe
general aspects of supersymmetric vacua of the $N=1$ theory on
general K\"ahler-Ricci soliton in the context of Morse theory.

\subsection{BPS Equations for Flat Domain Walls}

The flat background domain walls can be viewed as
\begin{equation}
ds^2 = a^2(u,
\tau)\,\eta_{\underline{\nu}\,\underline{\lambda}}\,dx^{\underline{\nu}}
dx^{\underline{\lambda}}-du^2 \;,\label{flatDWans}
\end{equation}
where $\underline{\nu}, \underline{\lambda}=0,1,2$,
$\eta_{\underline{\nu}\,\underline{\lambda}}$ is the metric on the
three dimensional Minkowskian spacetime $\lR^{1,2}$, and then, the
components of the corresponding Ricci tensor  of the metric
(\ref{flatDWans}) are given by
\begin{eqnarray}
R_{\underline{\lambda}\,\underline{\nu}} &=&  \left[ \left(
\frac{a'}{a} \right)' + 3 \left( \frac{a'}{a} \right)^2
  \right] a^2
  \eta_{\underline{\lambda}\,\underline{\nu}}\;,\nonumber\\
  R_{33} &=& -3 \left[ \left( \frac{a'}{a} \right)'
  +  \left( \frac{a'}{a} \right)^2 \right] \;, \label{flatRiccians}
\end{eqnarray}
and the Ricci scalar has the form
\begin{equation}
 R = 6 \left[ \left( \frac{a'}{a} \right)' + 2 \left( \frac{a'}{a} \right)^2
 \right] \;,\label{flatRans}
\end{equation}
where $a' \equiv \partial a/\partial u$. Here, $a(u, \tau)$ is the
warped factor taken to be $\tau$ dependent.\\
\indent In order to derive a set of equations that preserves
partially supersymmetry on the walls, we firstly have to consider
the supersymmetry transformation (\ref{susytr}) on the background
(\ref{flatDWans}). Setting $\psi_{\mu} = \chi^i = 0$, $z = z(u,
\tau)$, and correspondingly, solving the equations
$\delta\psi_{\mu} = 0$, $\delta\chi^i = 0$, it then results in
\cite{GZ}
\begin{eqnarray}
\frac{a'}{a}  &=& \pm \, {\mathcal{W}}(\tau) \;, \nonumber\\
 z^i{'} &=& \mp 2 g^{i\bar{j}}(\tau)
\bar{\partial}_{\bar{j}}{\mathcal{W}}(\tau) \; , \label{gfe}\\
\bar{z}^{\bar{i}}{'} &=& \mp 2 g^{j\bar{i}}(\tau)
\partial_j{\mathcal{W}}(\tau) \;,\nonumber
\end{eqnarray}
where ${\mathcal{W}}(\tau) \equiv e^{K(\tau)} \vert W(z)\vert$,
$K(\tau) \equiv K(z, \bar{z}; \tau)$ is the K\"ahler potential,
and $W(z)$ is a holomorphic superpotential. In this case, the
warped factor $a$ is monotonically decreasing related
 to the c-function in the holographic correspondence
\cite{DWc}. The second and the third equations in (\ref{gfe}) are
the BPS equations for flat domain walls describing supersymmetric
gradient flows. Another relevant supersymmetric
 flows for our analysis is the renormalization group flow given
  by the beta functions
\begin{eqnarray}
 \beta^i &\equiv& a \frac{\partial z^i}{\partial a} = - 2g^{i \bar{j}}(\tau)
  \frac{\bar{\partial}_{\bar{j}}
 {\mathcal{W}}}{\mathcal{W}} \;, \nonumber\\
\bar{\beta}^{\bar{i}} &\equiv& a \frac{\partial
\bar{z}^{\bar{i}}}{\partial a} =  - 2g^{ \bar{i}j}(\tau)
  \frac{\partial_j
 {\mathcal{W}}}{\mathcal{W}} \;, \label{flatbeta}
\end{eqnarray}
  describing the behavior of
 the couplings $(z^i, \bar{z}^{\bar{i}})$ with respect to the energy
 scale $a$ in the
 context of the conformal field theory (CFT)
 on the boundary $\lR^{1,2}$ \cite{CDGKL, DWc, CDKV}.\\
 \indent Furthermore, the scalar potential (\ref{V01}) can be cast
 into the form
\begin{equation}
 V(z,\bar{z}; \tau) = 4\, g^{i\bar{j}}(\tau)\, \partial_i {\mathcal{W}}\,
 \bar{\partial}_{\bar{j}} {\mathcal{W}}
 - \frac{3}{M^2_P}\, {\mathcal{W}}^2 \; ,
\label{V}
\end{equation}
and then, its first derivative with respect to $(z,\bar{z})$ is
\begin{eqnarray}
\partial_i V &=& 4\, g^{j\bar{k}}\, \nabla_i
\partial_j {\mathcal{W}}\, \bar{\partial}_{\bar{k}} {\mathcal{W}}+
4\, g^{j\bar{k}}\, \partial_j {\mathcal{W}}\,
\partial_i \bar{\partial}_{\bar{k}} {\mathcal{W}}
 - \frac{6}{M^2_P}\, {\mathcal{W}}\,\partial_i{\mathcal{W}}\; ,\nonumber\\
\partial_{\bar{i}} V &=& 4\,
g^{j\bar{k}}\,
\bar{\nabla}_{\bar{i}}\bar{\partial}_{\bar{k}}{\mathcal{W}}\,
\partial_j{\mathcal{W}}+ 4\, g^{j\bar{k}}\,
\bar{\partial}_{\bar{k}}{\mathcal{W}}\, \bar{\partial}_{\bar{i}}
\partial_j {\mathcal{W}}
 - \frac{6}{M^2_P}\,
 {\mathcal{W}}\,\bar{\partial}_{\bar{i}}{\mathcal{W}}\;
 ,\label{dV}
\end{eqnarray}
where $\nabla_i\partial_j {\mathcal{W}}= \partial_i \partial_j
{\mathcal{W}} - \Gamma^k_{ij}\partial_k {\mathcal{W}}$. All above
quantities are useful to study properties of supersymmetric
Lorentz invariant vacua in the next subsection \footnote{For the
rest of paper we mention Lorentz invariant vacuum (vacua) as
vacuum (vacua) or ground state.}.

\subsection{General Picture of Supersymmetric Vacua}

Let us first discuss general properties of vacua of the $N=1$
theory. A point $p_0 \equiv (z_0, \bar{z}_0)$ is a vacuum if
\begin{equation}
 \partial_i V (p_0) = \partial_{\bar{i}} V (p_0) = 0 \; .
\label{vacon}
\end{equation}
Supersymmetry further demands that $p_0$ defines a critical point
of the real function ${\mathcal{W}}(\tau)$, namely
\begin{equation}
 \partial_i {\mathcal{W}} (p_0) =
 \partial_{\bar{i}} {\mathcal{W}} (p_0) = 0 \; ,
\label{critcon}
\end{equation}
which can be regarded as a fixed point of the BPS equations in
(\ref{gfe}). At $p_0$, the scalar potential (\ref{V}) becomes
\begin{equation}
 V(p_0; \tau) =
 - \frac{3}{M^2_P}\, {\mathcal{W}}^2(p_0; \tau) \equiv
 - \frac{3}{M^2_P}\, {\mathcal{W}}^2_0 \; ,
\label{V0}
\end{equation}
which shows that the spacetime is AdS with negative cosmological
constant (or ${\mathcal{W}}_0 \ne 0$) and the warped factor
\begin{equation}
 a(u, \tau) = a_0(\tau)\, e^{\pm {\mathcal{W}}_0 u}  \;. \label{flatwarp}
\end{equation}
The Hessian matrix of the scalar potential (\ref{V}) evaluated at
$p_0$ is given by
\begin{eqnarray}
\partial_i \partial_j V(p_0; \tau)  &=& - \frac{1}{M^2_P}
{\mathcal{M}}_{ij}(p_0; \tau) \bar{L}(p_0; \tau) \;, \nonumber\\
\bar{\partial}_{\bar{i}} \bar{\partial}_{\bar{j}} V (p_0; \tau)
&=& - \frac{1}{M^2_P} \bar{{\mathcal{M}}}_{\bar{i}\bar{j}}(p_0;
\tau) L(p_0; \tau) \;, \label{HessV}\\
 \partial_i \bar{\partial}_{\bar{j}}V(p_0; \tau) &=&
 g^{k\bar{l}}(p_0; \tau) {\mathcal{M}}_{ik}(p_0; \tau)
 \bar{{\mathcal{M}}}_{\bar{l}\bar{j}}(p_0; \tau)
   - \frac{2}{M^4_P} g_{i\bar{j}}(p_0; \tau)
  {\mathcal{W}}_0^2  \:, \nonumber
\end{eqnarray}
where the quantities
\begin{eqnarray}
  L(p_0; \tau) &=& e^{K(p_0; \tau)/2M_P^2} W(z_0) \;, \nonumber\\
{\mathcal{M}}_{ij}(p_0; \tau) &=& e^{K(p_0; \tau)/2M_P^2} \left(
\partial_i  \partial_j W(z_0) + \frac{1}
{M_P^2}K_{ij}(p_0; \tau)W(z_0) + \frac{1}{M_P^2} K_j(p_0;
\tau)\partial_i W(z_0)\right), \nonumber\\
 && \label{fermass}
\end{eqnarray}
with their complex conjugate are related to the couplings in
two-fermions terms in the $N=1$ Lagrangian providing the masses of
the gravitino field and the masses of the spin-half fermions,
respectively. Moreover, using real coordinates defined as $z^i
\equiv x^i + {\mathrm{i}}x^{i+n}\,$ one can then show that the
Hessian matrix (\ref{HessV}) in this case is indeed a Hermitian
matrix, and therefore, it only admits real eigenvalues.\\
\indent The eigenvalues of the Hessian matrix (\ref{HessV})
determine the stability of domain walls in the context of
 dynamical system. For negative eigenvalues we have
unstable walls since the gradient flow provided by the last two
equations in (\ref{gfe}) is unstable along this direction, whereas
stable walls are for positive eigenvalues. In addition, the number
of negative eigenvalues of  (\ref{HessV}) is called the Morse
index of a vacuum.\\
\indent Then, the existence of the vacuum $p_0$ could also be
checked by using the first order expansion of the beta function
(\ref{flatbeta}) at $p_0$, namely
\begin{equation}
{\mathcal{U}} \equiv - \left(
\begin{array}{ccc}
  \partial_j \beta^i  & & \partial_j \bar{\beta}^{\bar{i}} \\
  & & \\
  \bar{\partial}_{\bar{j}} \beta^i  & &
  \bar{\partial}_{\bar{j}} \bar{\beta}^{\bar{i}} \\
\end{array}
\right)(p_0; \tau)\:, \label{flatbeta1}
\end{equation}
where
\begin{eqnarray}
\partial_j \beta^i(p_0; \tau) &=& - 2
g^{i \bar{k}}(p_0; \tau) \frac{\partial_j \bar{\partial}_{\bar{k}}
 {\mathcal{W}}_0}{{\mathcal{W}}_0} =
 -\frac{1}{M_P^2}\, \delta^i_j \;,\nonumber\\
\bar{\partial}_{\bar{j}} \beta^i(p_0; \tau) &=& - 2 g^{i
\bar{k}}(p_0; \tau) \frac{\bar{\partial}_{\bar{j}}
\bar{\partial}_{\bar{k}}
 {\mathcal{W}}_0}{{\mathcal{W}}_0} \;,
\end{eqnarray}
together with their complex conjugate with $\partial_j
\bar{\partial}_{\bar{k}}
 {\mathcal{W}}_0 \equiv \partial_j \bar{\partial}_{\bar{k}}
 {\mathcal{W}}(p_0; \tau)$. It is easy then to see that the matrix
 (\ref{flatbeta1}) has real eigenvalues since it is also a Hermitian
 matrix. Thus, we have a consistent theory.\\
\indent In the ultraviolet (UV) region we have the energy scale $a
\to +\infty$ and the matrix (\ref{flatbeta1}) must have at least a
positive eigenvalue because the RG flow is flowing away from the
region in this direction. On the other side, namely in the
infrared (IR) region, the energy scale $a \to 0$ and the RG flow
approaches the vacuum
in the direction of negative eigenvalue of (\ref{flatbeta1}).\\
\indent Here, we consider a general case, but the
 results are similar as in the K\"ahler-Einstein case
 \cite{GZ}. First of all, we consider a case where only nondegenerate
vacuum exists as follows.

\newtheorem{Vmorse}[Korbi]{Theorem}
\begin{Vmorse}\label{LemmaVmorse}
Let the scalar potential ${\mathcal{V}}(\tau)$ be a Morse
function, \textit{i.e.} no degenerate vacuum exists. Defining
\begin{eqnarray}
V_{ij}(p_0; \tau)  & \equiv &
 - \frac{\varepsilon(\sigma)}{M^2_P} \, {\mathcal{M}}_{ij}(p_0; \tau)
 \bar{L}(p_0; \tau) \;,\label{Vij}\\
 V_{i\bar{j}}(p_0;\tau) & \equiv &
 \vert g^{k\bar{l}}(p_0; \tau)\vert {\mathcal{M}}_{ik}(p_0; \tau)
 \bar{{\mathcal{M}}}_{\bar{l}\bar{j}}(p_0; \tau) \nonumber\\
   && - \frac{2}{M^4_P} \vert g_{i\bar{j}}(p_0; \tau) \vert
  {\mathcal{W}}_0^2 \:, \nonumber
\end{eqnarray}
where
\begin{equation}
\varepsilon(\sigma) =\left\{ \begin{array}{ll}
1 &   {\text{if}} \;\;\;  0 \le \tau < 1/2\Lambda \;, \\
-1 &  {\text{if}} \;\qquad \; \, \tau > 1/2\Lambda \;,\\
\end{array}
 \right.\label{varep}
 \end{equation}
 with $\Lambda >0$ for $\tau \ge 0$ and taking the inequalities
\begin{eqnarray}
{\mathrm{Re}}\big(V_{ij}(p_0; \tau) \big) > 0 \;, \quad
{\mathrm{Im}}\big(V_{ij}(p_0; \tau)
\big)> 0 \;, \nonumber\\
{\mathrm{Re}}\big(V_{i\bar{j}}(p_0; \tau)\big) > 0 \;, \quad
{\mathrm{Im}}\big(V_{i\bar{j}}(p_0; \tau)\big) > 0 \;,
\label{paritycon}
\end{eqnarray}
for $\tau \ge 0$ and $\tau \ne 1/2\Lambda$, then there exists a
parity transformation of the Hessian matrix (\ref{HessV}) caused
by the K\"ahler-Ricci soliton (\ref{KRsol}). In other words, if
$p_0$ is a vacuum of the index $\lambda$ in $\tau < 1/2\Lambda$,
then it becomes $\hat{p}_0$ of the index $2n- \lambda$ in $\tau >
1/2\Lambda$.
\end{Vmorse}

\noindent The proof of the above theorem has similar way with the
case of K\"ahler-Einstein geometries discussed in \cite{GZ}. We
have assumed that the K\"ahler-Ricci soliton is well defined for
$\tau \ge 0$ unless at $\tau = 1/2\Lambda$. Note that since
${\mathcal{V}}(\tau)$ is Morse function, then no index
modification of the index $\lambda$ in $\tau < 1/2\Lambda$ and of
the index $2n- \lambda$ in $\tau > 1/2\Lambda$. If the conditions
(\ref{paritycon}) do not hold, then there is no
parity transformation and the index remains unchanged.\\
\indent To make the above statements clearer, we now take a look
at a case where all spin-$\small{\frac{1}{2}}$ fermions are
massless at the ground states. This means
\begin{equation}
{\mathcal{M}}_{ij}(p_0; \tau) = 0 \quad \Rightarrow \quad
\partial_i \partial_j {\mathcal{W}}_0 = 0 \;.
\end{equation}
Then, the Hessian matrix (\ref{HessV}) simplifies to
\begin{equation}
\partial_i \bar{\partial}_{\bar{j}}V(p_0; \tau) =
    - \frac{2}{M^4_P} g_{i\bar{j}}(p_0; \tau)
  {\mathcal{W}}_0^2  \:, \label{HessV1}
\end{equation}
and the matrix (\ref{flatbeta1}) becomes
\begin{equation}
{\mathcal{U}} \equiv \frac{1}{M_P^2}\left(
\begin{array}{ccc}
  \delta^i_j   & & 0 \\
  & & \\
  0  & &
  \delta^{\bar{i}}_{\bar{j}} \\
\end{array}
\right)\:, \label{flatbeta2}
\end{equation}
showing that these vacua exist only in UV region. If
$g_{i\bar{j}}(p_0; \tau)$ is positive definite for finite $\tau$,
then the walls are unstable in this case. The other case, namely
for negative definite $g_{i\bar{j}}(p_0; \tau)$ we have only
stable walls. If $g_{i\bar{j}}(p_0; \tau)$ vanishes, then we have
singularity of the theory.\\
\indent Next, we consider a generalized case where the theory
possibly admits degenerate vacua. Or in other words, the scalar
potential (\ref{V}) is a Morse-Bott function. The construction can
be structured as follows. Let $S$ be a $m$ dimensional be a vacuum
submanifold of a K\"ahler geometry ${\bf M}$. Then at any $p_0 \in
S$ we can split the tangent space $T_{p_0} {\bf M} $ as
\begin{equation}
T_{p_0}{\bf M}  = T_{p_0}S \oplus N_{p_0}S  \;,
\end{equation}
where $T_{p_0}S$ is the tangent space of $S$ and $N_{p_0}S$ is the
normal space $S$. Moreover, the Hessian matrix (\ref{HessV}) is
non-degenerate in the normal direction to $S$. So we have a rich
and complicated structure of vacua. In the following we list some
possibilities.

\newtheorem{Modmorse}[Korbi]{Theorem}
\begin{Modmorse}\label{Modmorselemma}
Let $S$ be an $m$-dimensional submanifold of ${\bf M}$ with index
$\underline{\lambda}$
 in $0 \le \tau < \tau_0$ and
 $\tau_0 < 1/2\Lambda$. Then we have the following cases.

\begin{itemize}
\item[1.] $S$ deforms to an $m_1$-dimensional vacuum submanifold
$S_1 \subseteq M$ of the index $\underline{\lambda}_1$ in $\tau_0
\le \tau < 1/2\Lambda$. If $m_1 \ne m$, then the index
$\underline{\lambda}_1 \in \{0,...,2n - m_1 \}$. On the other
hand, if $m_1 =m$, then $\underline{\lambda}_1 \ne
\underline{\lambda}$ and $\underline{\lambda}_1 \in \{0,...,2n
-m_1 \}$.

\item[2.] If the inequalities (\ref{paritycon}) hold, then there
exists a parity pair of $S$, namely an $m$-dimensional submanifold
$\widehat{S} \subseteq M$ of the index $2n- \underline{\lambda}$
in $\tau > 1/2\Lambda$.

\item[3.] If the inequalities (\ref{paritycon}) do not hold, then
$S$ deforms to an $n_1$-dimensional vacuum submanifold
$\widehat{S}_1 \subseteq M$ of the index $\underline{\lambda}_2$
in $\tau
> 1/2\Lambda$.  If $n_1 \ne m$, then we have
$\underline{\lambda}_2 \in \{0,...,2n_c -n_1 \}$. However, if $n_1
=m$, then $\underline{\lambda}_2 \ne 2n -\underline{\lambda}$ and
$\underline{\lambda}_2 \in \{0,...,2n -n_1 \}$.
\end{itemize}
Furthermore, $\widehat{S}$ is the parity pair of $S$, and the
others, namely  $S_1$ and $\widehat{S}_1$, are not the parity
pair. All $S$, $S_1$, $\widehat{S}$, and $\widehat{S}_1$ may exist
in the UV or IR regions.
\end{Modmorse}

\noindent We leave the proof of Theorem \ref{Modmorselemma} in
appendix \ref{PT3}.\\
\indent We close this section by pointing out flat Minkowskian
vacua. In this case we have the condition
\begin{equation}
 \partial_i W (z_0) = W (z_0) = 0 \;,
\label{critconMink}
\end{equation}
which implies that the matrix (\ref{flatbeta1}) diverges.
Therefore, these vacua do not correspond to the AdS/CFT
correspondence. This case is excluded in the paper.

\section{Generalization to Curved BPS Domain Walls}
\label{curvedDW}
 In this section we generalize the previous case
to the curved (AdS sliced) domain walls. The structure of this
section is as follows. Firstly, we discuss shortly some aspects of
the curved domain walls on arbitrary dimensional K\"ahler-Ricci
soliton. These have been studied for two dimensional case in
\cite{GZ1}. Secondly, general properties of vacua will be
discussed.

\subsection{BPS Equations For Curved Domain Walls}

Similar as in the flat case, we consider the curved domain walls
by taking the ansatz metric of the four dimensional spacetime as
\begin{equation}
ds^2 = a^2(u,\tau)\,
g_{\underline{\lambda}\,\underline{\nu}}\,dx^{\underline{\lambda}}\,
dx^{\underline{\nu}} \,- du^2 \;,\label{curvedDWans}
\end{equation}
where $\underline{\lambda},\,\underline{\nu}=0,1,2$, $a(u, \tau)$
is again the warped factor, and
$g_{\underline{\lambda}\,\underline{\nu}}$ is the metric on the
three dimensional AdS spacetime. Therefore, the corresponding
components of the Ricci tensor  of the metric (\ref{curvedDWans})
are given by
\begin{eqnarray}
R_{\underline{\lambda}\,\underline{\nu}} &=&  \left[ \left(
\frac{a'}{a} \right)' + 3 \left( \frac{a'}{a} \right)^2
  - \frac{\Lambda_3 }{a^2}\right] a^2
  g_{\underline{\lambda}\,\underline{\nu}}\;,\nonumber\\
  R_{33} &=& -3 \left[ \left( \frac{a'}{a} \right)'
  +  \left( \frac{a'}{a} \right)^2 \right] \;, \label{curvedRiccians}
\end{eqnarray}
and the Ricci scalar has the form
\begin{equation}
 R = 6 \left[ \left( \frac{a'}{a} \right)' + 2 \left( \frac{a'}{a} \right)^2
 \right] - \frac{3 \Lambda_3 }{a^2}\;,\label{curvedRans}
\end{equation}
where $\Lambda_3$ is the
negative three dimensional cosmological constant.\\
\indent A set of equations that describes curved domain walls with
residual supersymmetry, can be derived by writing the
supersymmetry transformation (\ref{susytr}) on the background
(\ref{curvedDWans}). Using similar way as in the flat case, it
then results in \cite{GZ1}
\begin{eqnarray}
\frac{a'}{a}  &=& \pm \,\left\vert e^{K(\tau)/2M^2_P}\,  W(z) -
\frac{\ell}{a} \right\vert \;, \nonumber\\
 z^i{'}  &=& \mp  \,
2e^{{\mathrm{i}} \theta(\tau)}\,g^{i\bar{j}}(\tau)\,
\bar{\partial}_{\bar{j}}
{\mathcal{W}}(\tau) \;, \label{curvedgfe}\\
\bar{z}^{\bar{i}}{'}  &=& \mp \, 2 e^{-{\mathrm{i}}
\theta(\tau)}\,g^{\bar{i}j}(\tau)\,
\partial_j{\mathcal{W}}(\tau) \;, \nonumber
\end{eqnarray}
where  the phase function $\theta(z,\bar{z};u, \tau)$  has been
introduced with
\begin{equation}
e^{{\mathrm{i}} \theta(\tau)} = \frac{\left(1 - \ell
e^{-K(\tau)/2M^2_P}\, (aW)^{-1} \right)}{\left\vert 1 - \ell
e^{-K(\tau)/2M^2_P}\, (aW)^{-1} \right\vert} \;. \label{fase}
\end{equation}
Note that at $\theta =0$ the flat domain wall case is regained,
which corresponds to $\ell =0$. The second and the third equations
in (\ref{curvedgfe}) are called the BPS equations for curved
domain walls. Again, the renormalization group (RG) flow is given
by the beta functions
\begin{eqnarray}
 \beta_c^i(\tau) &\equiv& a \frac{\partial z^i}{\partial a} =
 - \frac{2e^{{\mathrm{i}} \theta(\tau)}}
 {\left\vert e^{K(\tau)/2M^2_P}\,  W(z) - \ell / a \right\vert}\,
 g^{i\bar{j}}(\tau)\, \bar{\partial}_{\bar{j}}
{\mathcal{W}}(\tau) \;,\nonumber\\
\bar{\beta}_c^{\bar{i}}(\tau) &\equiv& a \frac{\partial
\bar{z}^{\bar{i}}}{\partial a} = - \frac{2e^{-{\mathrm{i}}
\theta(\tau)}}{\left\vert e^{K(\tau)/2M^2_P}\, W(z) - \ell / a
\right\vert}\,g^{\bar{i}j}(\tau)\,
\partial_j{\mathcal{W}}(\tau) \;. \label{curvedbeta}
\end{eqnarray}
which give a description of a conformal field theory (CFT) on the
three dimensional AdS spacetime. Therefore, the scalars $(z^i,
\bar{z}^{\bar{i}})$ and the warped factor $a$ can be viewed as
coupling constants and  an energy scale, respectively \cite{CDGKL,
DWc, CDKV}.

\subsection{Generalized Vacuum Structure}
As mentioned in the preceding section, vacua of the $N=1$ theory
are in general defined by (\ref{vacon}). Then, in order to
maintain supersymmetry one has to add the condition
(\ref{critcon}) which can be viewed as the critical point of the
BPS equations in (\ref{curvedgfe}). The scalar potential (\ref{V})
evaluated at a vacuum $p_0$ has the form (\ref{V0}), however it is
not the cosmological constant of the four dimensional spacetime.\\
\indent In this case the warped factor at $p_0$ is given by
\begin{equation}
a(u, \tau)  = \frac{l}{{\mathcal{W}}^2_0} \pm \left(
\frac{\ell^2}{{\mathcal{W}}^2_0} -
\frac{l^2}{{\mathcal{W}}^4_0}\right)^{1/2} \left[A_0 \, e^{\pm
{\mathcal{W}}_0 u} - A^{-1}_0 e^{\mp {\mathcal{W}}_0 u}\right] \;,
\label{curvedwarp}
\end{equation}
where $ l \equiv \ell \, e^{K(p_0; \tau)/2M^2_P}
\,{\mathrm{Re}}W(z_0)$ and $A_0 \ne 0$. Since $a$ is real, then
${\mathcal{W}}_0 > \vert l \vert / \ell$. Moreover, we have
$\left(a'/a\right)' \ne 0$  near $p_0$. Inserting
(\ref{curvedwarp}) into (\ref{curvedRiccians}), one can then show
that the spacetime in general is non-Einstein whose components of
the Ricci tensor are given by
\begin{eqnarray}
R_{\underline{\lambda}\,\underline{\nu}} &=&  \left( \pm k' + 3
k^2 + 2\ell^2 \, e^{\mp 2 \int k \, du} \right) e^{\pm 2 \int k \,
du}  g_{\underline{\lambda}\,\underline{\nu}}\;,\nonumber\\
  R_{33} &=& -3 \left(\pm k' + k^2 \right)  \;, \label{curvedRiccians1}
\end{eqnarray}
where
\begin{equation}
k \equiv \left\vert e^{K(p_0; \tau)/2M^2_P}\,  W(z_0)
-\frac{\ell}{a} \right\vert  \;. \label{curvedk}
\end{equation}
 Note that
if we take $\ell \to 0$ and $A_0 \to \pm \infty$, then we obtain
the flat wall case \cite{GZ, GZA}. For $k=0$, one has
\begin{equation}
  {\mathrm{Im}}W(z_0) =0 \;, \label{singconbeta}
\end{equation}
and the spacetime becomes simply $AdS_3 \times \lR$ which
corresponds to the singularity of the beta function
(\ref{curvedbeta}). So, this vacuum is not related to the CFT on
$AdS_3$.\\
\indent Since the ground states are defined by (\ref{vacon}), the
next step would be to analyze the Hessian matrix of the scalar
potential given by (\ref{HessV}). These have been described by
Theorem \ref{LemmaVmorse} for only nondegenerate cases and Theorem
\ref{Modmorselemma} for general cases. Thus, this level is the
common step.\\
 \indent Next, similar as in the
flat case, we also need to look at the RG flows described by the
first order expansion of the beta function (\ref{curvedbeta}) at
$p_0$, namely
\begin{equation}
{\mathcal{U}}_c \equiv - \left(
\begin{array}{ccc}
  \partial_j \beta_c^i  & & \partial_j \bar{\beta}_c^{\bar{i}} \\
  & & \\
  \bar{\partial}_{\bar{j}} \beta_c^i  & &
  \bar{\partial}_{\bar{j}} \bar{\beta}_c^{\bar{i}} \\
\end{array}
\right)(p_0; \tau)\:, \label{curvedbeta1}
\end{equation}
where
\begin{eqnarray}
\partial_j \beta_c^i(p_0; \tau) &=&
 -\frac{e^{{\mathrm{i}}\theta_0(\tau)}}{k \,M_P^2}\, {\mathcal{W}}_0 \,
 \delta^i_j \;,\nonumber\\
\bar{\partial}_{\bar{j}} \beta_c^i(p_0; \tau) &=& -
\frac{2e^{{\mathrm{i}}\theta_0(\tau)}}{k} \, g^{i \bar{k}}(p_0;
\tau) \bar{\partial}_{\bar{j}} \bar{\partial}_{\bar{k}}
 {\mathcal{W}}_0  \;,
\end{eqnarray}
together with their complex conjugate with $\theta_0(\tau) \equiv
\theta(p_0; \tau)$. In general, the matrix (\ref{curvedbeta1}) is
not Hermitian and thus, has complex eigenvalues. In order to have
a consistent theory, one has to diagonalize (\ref{curvedbeta1})
and then, in this basis imposes Hermiticity condition on it. This
will result a consistency condition for RG flows.\\
\indent In two dimensional cases, it is much easier to achieve the
consistency condition in which the eigenvalues of
(\ref{curvedbeta1}) are real. As studied in \cite{GZ1}, such
condition does exist, namely
\begin{equation}
 \vert {\mathrm{Tr}} {\mathcal{U}}_c \vert \ge
 2 \, \vert {\mathrm{Det}} {\mathcal{U}}_c \vert^{1/2}  \;.
\end{equation}

\section{Vacuum Structure Near Orbifold Point}
\label{VSNO}

In this section, some aspects of the vacua on the K\"ahler-Ricci
orbifold are discussed. We particularly consider those aspects
near the origin (which is the orbifold point), since around $\xi
\to +\infty$ the soliton becomes the K\"ahler-Einstein geometry
$\,{\mathrm{\lC P}}^{n -1}$ which has been studied in \cite{GZ}.\\
 \indent Before turning to the main discussion of this
section, let us first mention two quantities which are useful for
our analysis around the orbifold point, namely the origin. These
are the $U(1)$ connection and the K\"ahler potential related to
the metric (\ref{nearzerosoliton}) of the theory given by
\begin{eqnarray}
 Q(\xi,\bar{\xi};\tau) &=&  -\frac{2\mathrm{i}}{M_P^2} \,
 \sigma(\tau)^{1-\mu /\Lambda} \,
 B\left(\sigma(\tau)^{-\mu /\Lambda} \vert \xi \vert^2 \right)
 \left[ \bar{\xi}^{\bar{i}} \, d\xi^i - \xi^i \, d\bar{\xi}^{\bar{i}}
 \right]\;, \nonumber\\
K(\xi,\bar{\xi};\tau) &=& \sigma(\tau)^{1-\mu /\Lambda} \int
e^{v/2} B\left(\sigma(\tau)^{-\mu /\Lambda}e^{ v/2} \right) dv + c
\;, \label{U1conex}
\end{eqnarray}
respectively where $v \equiv 2\,{\mathrm{ln}} \vert \xi \vert^2$
 and $c$ is a real constant. Here, the order $1-\mu /\Lambda =
 p/q$ satisfies the conditions given in Theorem \ref{Korbi}.\\
 \indent Firstly, we consider the case where the scalar potential
 (\ref{V}) is Morse function. Let $p_0 \equiv (\xi_0,\bar{\xi}_0)$
 be a ground state and exists around the origin. We assume that
 $p_0 \ne 0$ and it is not an orbifold point. Then, for the
 cases 1 and 2 of Theorem \ref{Korbi},
 it might be possible to have a situation described in Theorem
 \ref{LemmaVmorse} if the condition (\ref{paritycon}) is fulfilled.
  In the massless case, the case 1 can be easily seen, but
  it might be not possible for the case 2 if the sign of both
  $B\left(\sigma(\tau)^{-\mu /\Lambda} \vert \xi \vert^2 \right)$
  and $\dot{B}\left(\sigma(\tau)^{-\mu /\Lambda} \vert \xi \vert^2
  \right)$ could not be altered after $\tau = 1/2\Lambda$. For the
  case 3 of Theorem \ref{Korbi} and Lemma \ref{Korbi1}, the
  condition (\ref{paritycon}) does not exist since there is
  no K\"ahler geometry in $\tau > 1/2\Lambda$.\\
\indent Secondly, if the scalar potential (\ref{V}) is Morse-Bott
function, then we might have the situations stated in Theorem
\ref{Modmorselemma} for the cases 1 and 2 of Theorem \ref{Korbi}.
However, only the situation 1 in Theorem \ref{Modmorselemma} are
possible for the case 3 of Theorem \ref{Korbi} and Lemma
\ref{Korbi1} because no K\"ahler
geometry exists after the singularity at $\tau = 1/2\Lambda$.\\
\indent In the following we simply take a simple case near $\tau
\to + \infty$. So, it is valid only for the cases 1 and 2 of
Theorem \ref{Korbi}. Since $\xi \approx 0$, then the
 quantities in (\ref{U1conex}) are simplified to
\begin{eqnarray}
 Q(\xi,\bar{\xi};\tau) &\approx&  -\frac{2\mathrm{i}}{M_P^2} \,
 \sigma(\tau)^{p/q} \,
 B\left(0 \right)
 \left[ \bar{\xi}^{\bar{i}} \, d\xi^i - \xi^i \, d\bar{\xi}^{\bar{i}}
 \right]\;, \nonumber\\
K(\xi,\bar{\xi};\tau) &\approx& 2 \sigma(\tau)^{p/q} \, B\left(0
\right) \vert \xi \vert^2 + c \; . \label{U1conexapp}
\end{eqnarray}
Moreover, for a sake of simplicity the form of the superpotential
$W(\xi)$ is taken to be linear, namely
\begin{equation}
 W(\xi) = a_0 + a_i \, \xi^i \;,
\label{linearsup}
\end{equation}
with $a_0, a_i \in \lR$. For $a_i =0$ and $a_0 \ne 0$, we find
that the solution of (\ref{critcon}) is the origin which is the
orbifold point. This further implies that all
spin-$\small{\frac{1}{2}}$ fermions are massless, but it is an ill
defined $N=1$ theory. On the other case, for finite $a_i > 0$,
$a_0 > 0$, and $a_0 \gg a_i$, the equation (\ref{critcon}) gives
\begin{equation}
\xi^i_0(\tau)  \approx - \frac{1}{2} \, M_p^2 \,
\sigma(\tau)^{-p/q} \, \frac{a_i}{a_0} \;, \label{vacsol}
\end{equation}
which can then be shown that all spin-$\small{\frac{1}{2}}$
fermions are massive. In this case we may have a well defined
theory.

\section{Conclusions}
\label{conclu}

First of all, we particularly have studied the $U(n)$
K\"ahler-Ricci soliton that admits an orbifold-type singularity at
the origin. This soliton may be considered as a one-parameter
family of K\"ahler geometries. As mentioned in Theorem
\ref{Korbi}, near the orbifold point (or the origin) if there
exists a rational number $p/q = 1 - \mu/\Lambda$ with $0 < p < q$,
then there are three possible situations. In the situations 1 and
2, we possibly have a family of K\"ahler geometries even after the
singularity at $\tau = 1/2\Lambda$. For the situation 3, the
K\"ahler geometry exists only in $\tau < 1/2\Lambda$. Such
situation also occurs in Lemma  \ref{Korbi1}.\\
\indent Next, some aspects of the $N=1$ supergravity
domain walls on K\"ahler-Ricci soliton have been discussed.
Firstly, we considered flat domain walls together with their
vacuum structure. For the case at hand, it is natural that both
the Hessian matrix (\ref{HessV}) and the matrix (\ref{flatbeta1})
describing RG flows are Hermitian matrices and therefore, they
have real eigenvalues. Thus, we have a consistent theory for the
flat case. In addition, a ground state is characterized by the
pair $(m, \lambda)$ where $m$ and $\lambda$ are the dimension and
the Morse index of a ground state, respectively. As stated in
Theorems \ref{LemmaVmorse} and \ref{Modmorselemma}, the pair $(m,
\lambda)$ may in general be deformed with respect to the flow
parameter of the K\"ahler-Ricci soliton. The results here is the
same as in the previous results for K\"ahler-Einstein geometry
\cite{GZ}.\\
\indent Secondly, curved domain walls with their vacuum structure
have also been discussed. In this case, the pair $(m, \lambda)$
also characterizes a vacuum and possibly has a deformation caused
by the K\"ahler-Ricci soliton. However, the matrix
(\ref{curvedbeta1}) describing RG flows has in general complex
eigenvalues. Then, we have to impose Hermiticity condition in the
basis where the matrix (\ref{curvedbeta1}) has the diagonal form.\\
\indent Finally, we have considered some aspects of vacuum
structure on the $U(n)$ K\"ahler-Ricci orbifold near the origin.
The cases 1 and 2 of Theorem \ref{Korbi}  have a possibility of
having the situations described in Theorems \ref{LemmaVmorse} and
\ref{Modmorselemma}. But in the case 3 of Theorem \ref{Korbi} and
Lemma \ref{Korbi1}, the possible case is only the case 1  of
Theorem \ref{Modmorselemma}. At the end, we simply considered a
simple case when the flow parameter $\tau \to + \infty$ and the
superpotential $W(\xi) = a_0 + a_i \, \xi^i$. We find that the
vacuum structure related to $a_0 \ne 0$ and $a_i =0$ is an ill
defined massless $N=1$ theory at the origin (which is the orbifold
point). On the other side, for finite and positive definite $a_0,
\, a_i$ and $a_0 \gg a_i$ we may have a well defined theory which
is valid only for the cases 1 and 2 of Theorem \ref{Korbi}.

\vskip 1truecm

\hspace{-0.2 cm}{\Large \bf Acknowledgement}
\\
\vskip 0.15truecm \hspace{-0.6 cm} \noindent We thank L.
Andrianopoli, J. M. Maldacena, T. Mohaupt,  and M. Zagermann for
useful discussions. One of us (BEG) is grateful to the organizer
of the Abdus Salam ICTP Spring School on Superstring Theory and
Related Topics for providing a good atmosphere where parts of this
work was done. This work was initiated  by Hibah Kompetensi DIKTI
2009 No. 223/SP2H/PP/DP2M/V/2009 and Riset KK ITB 2009 No.
243/K01.7/PL/2009, and is extensively supported by ITB Alumni
Association Research Grant (HR IA-ITB) 2009 No. 180a/K01.7/PL/2009
and Hibah Kompetensi DIKTI 2010 no. 227/SP2H/PP/DP2M/III/2010.

\appendix

\section{Proof of Theorem \ref{Korbi} and Lemma \ref{Korbi1}}
\label{PT2}

Let us first look at the functions $B(\vert \xi \vert^2)$ and
$\dot{B}(\vert \xi \vert^2)$ defined in (\ref{nearzerometric}).
Since $\xi \approx 0$ those functions can be expanded as
\begin{eqnarray}
 B(\vert \xi \vert^2) &=& B(0) + \vert \xi \vert^2 \dot{B}(0)
 + O(\xi, \bar{\xi})\;, \nonumber\\
\dot{B}(\vert \xi \vert^2) &=& \dot{B}(0) + \vert  \xi \vert^2
\ddot{B}(0) + \widetilde{O}(\xi, \bar{\xi})\; . \label{Bexpand}
\end{eqnarray}
Positivity of the metric (\ref{nearzerometric}) implies
$\dot{B}(\vert \xi \vert^2) > B(\vert \xi \vert^2)$. Employing the
diffeomorphism (\ref{diffeo2}) we obtain
\begin{eqnarray}
 B(\sigma(\tau)^{-\mu /\Lambda} \vert \xi \vert^2) &=&
 B(0) + \sigma(\tau)^{-\mu /\Lambda} \vert \xi \vert^2 \dot{B}(0)
 + O(\xi, \bar{\xi}; \tau)\;, \nonumber\\
\dot{B}(\sigma(\tau)^{-\mu /\Lambda} \vert \xi \vert^2) &=&
\dot{B}(0) + \sigma(\tau)^{-\mu /\Lambda}\vert  \xi \vert^2
\ddot{B}(0) + \widetilde{O}(\xi, \bar{\xi}; \tau)\; .
\label{Bexpand1}
\end{eqnarray}
Thus, we have a consistent theory if the above functions are real
valued. Assuming that there exists a rational number $p/q = 1 -
\mu/\Lambda$ and $0 < p <q$. For the case 3 there are no K\"ahler
manifolds for $\tau > 1/2\Lambda$ since the functions
(\ref{Bexpand1}) become complex valued. In the case 1 the
functions (\ref{Bexpand1}) are always positive definite for $\tau
\ge 0$ and $\tau \ne 1/2\Lambda$ because the order $\mu/\Lambda$
has the form $2n/2m+1$ for some positive integer $m,n$. In the
case 2 for $\tau \to +\infty$ the functions (\ref{Bexpand1}) are
positive definite since the leading terms are $B(0)$ and
$\dot{B}(0)$. However, around $\tau = 1/2\Lambda$ the second and
the higher order terms would be dominant and the sign of the
functions (\ref{Bexpand1}) could be changed after $\tau =
1/2\Lambda$.\\
\indent Using the same statement as the case 3 of Theorem
\ref{Korbi}, we have thus proved Lemma \ref{Korbi1}.

\section{Proof of Theorem \ref{Modmorselemma}}
\label{PT3}

Our starting point is to consider the Hessian matrix of the scalar
potential $H_V$ whose components are given in (\ref{HessV}). In
particular, $H_V$ is a $2n \times 2n$ matrix in real coordinates
such that $z^i \equiv x^i + {\mathrm{i}}x^{n+i}$. First of all, we
want to mention that the proof of the case 2 is similar like in
Theorem \ref{LemmaVmorse} which has been proved for
K\"ahler-Einstein geometries. Generalization to any K\"ahler-Ricci
soliton is straightforward. So, the rest step is only to prove
that the cases 1 and 3 are possible. As we will see, the proof of
the case 1 has a similar logical way as the case 3. Therefore, we
just have to prove the case 1.\\
\indent Suppose at $\tau = \tau_0 < 1/2\Lambda$, we have an
$m$-dimensional submanifold $S$ of the index $\underline{\lambda}
= 2n-m$. This corresponds to the existence of $m$ numbers zero
eigenvalues and of $2n-m$ numbers negative eigenvalues of $H_V$.
Let us denote each eigenvalue by $\alpha_i$, and in this case
there are some $\alpha_i < 0, \;\, i=1,...,2n-m$.

\begin{itemize}
\item[i.] Then, at $\tau = \tau_1$ with $\tau_0 <
\tau_1<1/2\Lambda$ one of the eigenvalues vanishes, say
$\alpha_{2n-m}=0$. Thus, $S$ deforms to another $m+1$-dimensional
submanifold $S_1$ of the index $\underline{\lambda}_1 = 2n-m-1$.
This proves $m_1 \ne m$ and $\underline{\lambda}_1 \ne
\underline{\lambda}$.

\item[ii.] Then, at $\tau = \tau_2$ with $\tau_1 <
\tau_2<1/2\Lambda$ one of the eigenvalues becomes positive, say
$\alpha_{2n-m} > 0$. So, $S$ deforms to another $m$-dimensional
submanifold $S_1$ of the index $\underline{\lambda}_1 = 2n-m-1$.
This proves $m_1 = m$ and $\underline{\lambda}_1 \ne
\underline{\lambda}$.

\item[iii.] Comparing i. and ii., we have proved $m_1 \ne m$ and
$\lambda_1 = \lambda$.
\end{itemize}

The case 3 can be straightforwardly proved using the above steps.

\end{document}